\documentclass[twocolumn,showpacs,english,amsmath,amssymb, pra, aps, 10pt] {revtex4-1}

\usepackage{tgtermes}
\usepackage[utf8]{inputenc}
\usepackage{graphicx}
\usepackage{dcolumn}
\usepackage{bm}
\usepackage{xcolor}
\usepackage{makeidx}
\usepackage{amsfonts}
\usepackage{subfigure}
\usepackage{epsfig}
\usepackage{pst-grad} 
\usepackage{pst-plot} 
\usepackage[bookmarks=false,colorlinks=true, breaklinks=true]{hyperref}
\hypersetup
	{ colorlinks,%
		citecolor=green,%
		linkcolor=red,%
		filecolor=magenta,%
		urlcolor = cyan,
	}

\newcommand{\nn}{\nonumber\\}

\newcommand{\bea}{\begin{eqnarray}}
\newcommand{\ea}{\end{eqnarray}}
\newcommand{\eea}{\end{eqnarray}}

\newcommand{\sumint}[1]

\begin{document}

\title{ Spin-orbit-coupling-induced magnetic heterostructure in the bilayer Bose-Hubbard system}

\author{Bo Xiong,$^{1, 2}$ Jun-hui Zheng,$^{3}$ Yu-Ju Lin,$^{4}$ and Daw-wei Wang$^{2,3}$}
\affiliation{$^{1}$School of Science, Wuhan University of Technology, Wuhan 430070, China \\
             $^{2}$Institute of Physics, National Center for Theoretical Science, Hsinchu, Taiwan 300, Republic of China \\
             $^{3}$Department of Physics, National Tsing-Hua University, Hsinchu, Taiwan 300, Republic of China \\
             $^{4}$Institute of Atomic and Molecular Sciences, Academia Sinica, Taipei 10617, Taiwan
             }

\date{\today}

\begin{abstract}
   We investigate magnetic phase in the bilayer system of ultra-cold bosons in an optical lattice, which is involved with Raman-induced spin-orbit (SO) coupling and laser-assisted interlayer tunneling. It is shown that there exit a rich of spin textures such as hetero ferromagnet, heterochiral magnet, chiral magnet with interlayer antiferromagnet. In particular, heterochiral magnet induced by SO coupling occurs extremely rarely in real solid-state materials. We present detailed experimental setup of realizing such a model in cold atom system.  
   
\end{abstract}


\maketitle
\section{Introduction}  
   The spin-orbit (SO) coupling in the systems of ultracold bosonic atoms \cite{lin1} and fermionic atoms \cite{wang, cheu} have been realized experimentally. The combination of the creation technique of SO coupling with Feshbach resonances and optical lattice (OL) enables the production of exotic many-body states, e.g., the novel quantum spin states \cite{radi, cole} and the spin-Hall effect \cite{sino, kane}, which may be found in the conventional solid-state system but are difficult to be realized experimentally. Furthermore, in solid-state materials, the SO coupling stems from its essential parameters so that some relevant physics is hardly explored due to the limited system parameters. By contrast, artificial SO coupling in the cold atomic system can be adjusted in a large parameter regime via laser beams, which enable it a desirable candidate for exploiting some complex mechanism induced by the coupling.

   Recently, an increasing interest concerns on synthetic two-dimensional (2D) and higher-dimensional SO coupling, which is crucial for realizing high-dimensional topological matters. Several theoretical works propose some feasible ways to generate high-dimensional SO coupling in ultracold atomic system, for example, by employing optical field to couple internal states  \cite{ruse, juze1, juze2, camp, liu} or by using magnetic-field-gradient pulse \cite{ande, xu1} as well as Raman laser pulse \cite{xu2}. Since 1D SO coupling with an equal weight of Rashba and Dresselhaus terms has been achieved experimentally \cite{lin1, wang, cheu}, theoretical attempt to realize 2D (3D) SO coupling by employing 1D SO coupling in some special geometric systems has been presented; for example, analogous Rashba SO can be realized by 1D SO acting on the bilayer system despite the configuration of several lasers keeps experimental challenge \cite{sun}. Recent experiments for realizing 2D SO coupling in ultracold Fermi gases confirms the possibility of such routine \cite{huang}. 
   
   In what follows, we propose a theoretic model corresponding to an experimentally accessible bilayer system with 1D SO coupling to realize 2D SO coupling \cite{note1}. By employing optical lattice and taking strong interatomic interaction into account, we derive an effective spin Hamiltonian based on the second-order perturbation theory in Sec. II B. Moreover, we explore magnetic phases of such Hamiltonian and its origination associated with the SO coupling. We find that there exit a rich of spin textures such as hetero ferromagnet, heterochiral magnet, chiral magnet with interlayer antiferromagnet. In particular, heterochiral magnet induced by SO coupling occurs extremely rarely in real solid-state materials. Finally, we stress that our results provide qualitative understanding for complex magnetic phenomena, which is involved with interlayer tunneling and SO coupling, and the underlying mechanism might be relevant and inspired to some novel magnetic materials.

\section{Bilayer BEC in optical lattice} 
\label{secII}

\subsection{Single-particle Hamiltonian}

\begin{figure}[bt]
\centering
\includegraphics[scale = 0.51]{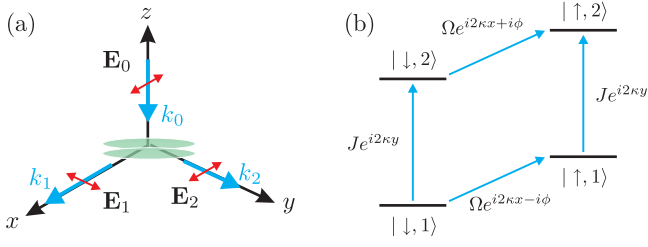} 		        	
\caption{(Color online) (a) Illustration of a specific laser configuration for producing Raman transitions and interlayer tunneling. All laser beams are polarized linearly: ${\bf E}_{0}$, ${\bf E}_{1}$, and ${\bf E}_{2}$ propagate along $\hat{z}$, $\hat{x}$, $\hat{y}$ direction and polarized along the $\hat{x}$, $\hat{y}$, and $\hat{x}$, respectively. (b) A level scheme involving two atomic internal ($ \gamma = \uparrow,\downarrow $) states and two layers $ a = 1, 2 $ formed by a double-well potential along the $z$-direction. The two-dimensional optical lattice, which is not shown in Fig.\,\ref{Fig1} (a), is switched on in the $ x y $ plane and the laser directions are respectively $x$- and $y$-axises. The atoms can move within each layer in the $x y$ plane and the layers are separated between each other by a distance $d$ in the $z$ direction. In a proper chosen frequencies, the Raman transition is realized by ${\bf E}_{0}$ and ${\bf E}_{1}$ whereas the laser-assisted interlayer tunneling is produced by ${\bf E}_{0}$ and ${\bf E}_{2}$. The phase difference $2\phi$ for the Raman coupling in two layers is contributed by ${\bf E}_{0}$ as it provides the $z$ component to the Raman coupling. We are particularly interested in the $\pi$ phase difference between layers, i.e., $\phi = \pi/2$. The more details have been shown in the Appendix \ref{ALI}.     
}
\label{Fig1}
\end{figure}

The bilayer system is fabricated by an atomic BEC with two internal (quasi-spin) states trapped in a tilted double-well potential along $z$ direction and 2D optical lattice in the $xy$ plane. In the initial step, we take the four combined spin-layer states $|\gamma, a\rangle \equiv |\gamma\rangle_{\rm spin} \otimes |a\rangle_{\rm layer}$ as the states required for the ring coupling scheme for generating the Rashba-type SOC \cite{camp}. Here $\gamma = \uparrow, \downarrow$ denotes an internal (quasi-spin) atomic state and $a = 1, 2$ labels the $a$-th layer. The level scheme for the single-particle tunneling process and the associated laser configuration are shown in Fig.\,\ref{Fig1}. The details for producing laser-assisted interlayer tunneling and intralayer SO coupling are given in Appendix \ref{ALI}. The relevant single-particle Hamiltonian yields, 
\bea
   \hat{H}_{0}^{'} = \frac{\hat{\bf P}^{2}}{2m} + V(\hat{\bf r}) + \hat{H}_{\rm inter}^{'} + \hat{H}_{\rm intra}^{'}
\ea
where the interlayer coupling term 
\bea
   \hat{H}_{\rm inter}^{'} = \sum_{\gamma = \uparrow, \downarrow} \left[J e^{i 2\kappa y} |\gamma, 2 \rangle \langle \gamma, 1| + {\rm h.c.}\right] \nonumber, 
\ea
and the intralayer coupling term
\bea
   \hat{H}_{\rm intra}^{'} = \Omega e^{i2 \kappa x} \left(e^{-i\phi} |\uparrow, 1\rangle \langle \downarrow, 1| + e^{i \phi} |\uparrow, 2 \rangle \langle \downarrow, 2| \right) + {\rm h.c.} \nonumber.
\ea   
Under the basis $|\psi\rangle = (|\psi_{2\uparrow}\rangle, |\psi_{2\downarrow}\rangle, |\psi_{1\uparrow}\rangle, |\psi_{1\downarrow}\rangle)$, we expand $\hat{H}_{0}^{'}$ into a matrix form and eliminate the position-dependent term in off-diagonal part by employing unitary transformation 
$\hat{H}_{0} = \hat{U}^{\dag} \hat{H}_{0}^{'} \hat{U}$ with
\bea
   \hat{U} = \begin{bmatrix} e^{-i\kappa(x + y)} & 0 & 0 & 0 \\
                             0 & e^{i\kappa(x-y)} & 0 & 0   \\
                             0 & 0 & e^{-i\kappa(x-y)} & 0   \\
                             0 & 0 & 0 & e^{i\kappa(x+y)}  \end{bmatrix}.
\ea

In the situation, we concentrate particularly on, that the $\pi$ phase difference induced by Raman coupling between the upper and lower layer ($\phi = \pi/2$), $\hat{H}_{0}$ reads in the second quantization terminology, 
\bea
   \hat{H}_{0} & = & {\bf \int} \Big\{\sum_{a} \left[ \hat{\psi}_{a}^{\dag} ({\bf r}) \epsilon_{a} \hat{\psi}_{a}({\bf r}) + (-1)^{a} \Omega \hat{\psi}_{a}^{\dag} ({\bf r}) \sigma_{y} \hat{\psi}_{a} ({\bf r}) \right]  \nn
               & + & J\hat{\psi}_{2}^{\dag} ({\bf r}) \hat{\psi}_{1}({\bf r}) + {\rm h.c.} \Big\} d{\bf r} \label{Hamiltonian2},
\ea
where $\sigma_{x, y, z}$ are the spin-$1/2$ representation of Pauli matrices, and $\epsilon_{a}({\bf r}) = \frac{[-i\nabla + {\bf A}_{a}]}{2m} + V({\bf r})$ with ${\bf A}_{a} = \kappa \hat{x} \sigma_{z} + (-1)^{a} \kappa \hat{y}$.

In the following step, we consider the influence of optical lattice on such system under the tight-binding approximation, where $\hat{\psi}_{a} ({\bf r}) = \sum_{i} \hat{\psi}_{a, i} \omega_{i}({\bf r})$. Thus in such a lattice system and considering the occupation of atoms in the lowest energy band, Hamiltonian (\ref{Hamiltonian2}) can be written into
\bea
   \hat{H}_{t} & = & - \sum_{a, i, \delta = \hat{x}, \hat{y}} t\left[ \hat{\psi}_{a,i}^{\dagger
}R_{a\delta}\hat{\psi}_{a, i + \delta} + {\rm h.c.} \right] \nn
         & + &   \sum_{a, i}\left( -1\right)^{a}\Omega \hat{\psi}_{a,i}^{\dagger }\sigma
_{y}\hat{\psi}_{a, i} + \sum_{i}\left[ J\hat{\psi}_{2,i}^{\dagger }\hat{\psi}_{1,i} + {\rm h.c.} \right], \label{hamiltonian3}
\ea
where $R_{ax} = \exp \left(i\mathbf{A}_{a}\cdot d \hat{x}\right)$
and $R_{ay} = \exp \left(i\mathbf{A}_{a}\cdot d \hat{y}\right)$. Here we have taken the grid size $d$ as the unit of length, i.e., $d = 1$, and $t=\int d {\bf r} \omega _{i}\left( {\bf r} \right) %
\left[ \frac{-\nabla ^{2}}{2m} + V\left( \mathbf{r}\right) \right] \omega
_{j}\left( \mathbf{r}\right)$. In Appendix \ref{SPD}, we show that the lowest energy dispersion according to Hamiltonian (\ref{hamiltonian3}) displays four degenerate minimal, in an analogous way to the 2D square lattice system with SO coupling \cite{cole, radi}. Since some prior works have discussed the (contact) interaction effect on the superposition of four degenerate minima and predicted some various superfluid pattern phases \cite{cole, sun, su}, we will not repeat such issue here and will concentrate on the strongly coupling Mott insulator regime where two-body interaction is much larger than all tunneling terms.

\subsection{Strongly coupling Mott regime}

   Without loss of generality, we consider that the two-body interaction yields the form, 
   \bea
      \hat{H}_{\rm int} & = & \frac{U}{2} \sum_{a, i} \left[\hat{n}_{a\uparrow, i} (\hat{n}_{a\uparrow, i} - 1) + \hat{n}_{a\downarrow, i} (\hat{n}_{a\downarrow, i} - 1) \right. \nn
                  & + & \left. 2 \lambda \hat{n}_{a \uparrow, i} \hat{n}_{a\downarrow, i} \right]
   \ea
where $U$ is onsite interatomic interaction coupling length and $\lambda$ is the ratio between the onsite interatomic interaction in the same internal state and the interaction in different internal states.       
   
   At unit filling and when $U/t$, $U/J$, $U/\Omega$ approaches infinity, the strong repulsive interactions restrict one boson occupy one site. In absence of any tunnelings, the spin states of the bilayer system are highly degenerate and an arbitrary spin state at each site can exist. When finitely large tunnelings are involved (but still much smaller than $U$), the degeneracy of the spin states is broken and some spin correlations and magnetic orders are built. Therefore we address specifically some exotic magnetic phenomena and its associated mechanism induced by these tunnelings in the bilayer system.    
   
   The already existing tunnelings, $t$, $J$, $\Omega$, enable the possibility of the transition between the ground state for $\Omega = t = J = 0$ and its exciting states. We denote the ground state and potential excitation state as $|{\rm GS}\rangle$ and $|{\rm ES} \rangle$, respectively.   
By employing perturbation theory up to $\mathcal{O}(t^{2}/U)$ \cite{note2}, we desire for deriving an effective spin Hamiltonian through 
   \bea
      \langle {\rm GS}| H_{\rm spin} |{\rm GS}\rangle & = & \langle {\rm GS}| \hat{H}_{\rm int} | {\rm GS} \rangle \nn
                                                     & - &\langle {\rm GS}| \hat{H}_{t} |{\rm ES}\rangle \langle {\rm ES} |
      \frac{1}{\hat{H}_{\rm int}} |{\rm ES} \rangle \langle {\rm ES} |\hat{H}_{t} |{\rm GS} \rangle. \nonumber
   \ea    
After some nontrivial treatment, we obtain the effective spin Hamiltonian in the strongly coupling Mott regime,
   \bea
       &&H_{\rm spin}   =  \sum_{a, i, \delta = \hat{x}, \hat{y}} \sum_{\alpha = x, y, z} J_{\delta}^{\alpha} S_{a, i}^{\alpha} S_{a, i + \delta}^{\alpha} 
                       + \sum_{a, i} 2 (-1)^{a} \Omega S_{a, i}^{y} \nn
                     & &~~+  \sum_{a, i} \vec{D}_{z} \cdot \left( \vec{S}_{a, i} \times \vec{S}_{a, i + \hat{x}} \right) + \sum_{i, \alpha = x, y, z} J_{\bot}^{\alpha} S_{1, i}^{\alpha} S_{2, i}^{\alpha},  \label{MI_eq1}
   \ea
where the inlayer exchange coupling constants $J_{\delta}^{\alpha}$ and the interlayer exchange coupling constants $J_{\bot}^{\alpha}$ have the following forms: $J_{\hat{x}}^{x} = J_{\hat{x}}^{y} = - \frac{4 t^{2} \cos (2 \kappa)} {U \lambda}$, $J_{\hat{y}}^{x} = J_{\hat{y}}^{y} = -\frac{4 t^{2}}{U\lambda}$, $J_{\hat{x}}^{z} = J_{\hat{y}}^{z} = - \frac{4t^{2}}{U} \left( 2 - \frac{1}{\lambda} \right)$, $J_{\bot}^{x} = J_{\bot}^{y} = - \frac{4 J^{2}} {U \lambda}$, and $J_{\bot}^{z} = - \frac{4 J^{2}}{U\lambda} \left(2 - \frac{1}{\lambda} \right)$. The third term is the so-called Dzyaloshinskii-Moriya (DM) term \cite{dzya, mori} and the DM vector $\vec{D}_{z} = -\frac{4 t^{2} \sin (2 \kappa)}{U \lambda} \hat{z}$. Owing to spin inversion symmetry breaking as a result of some topological magnetic orders, the DM interaction attracts much attention in the material science. Some rich spin textures, such as spin spirals \cite{bode, heid} and skyrmions \cite{yu, hein, huang1}, and stable chiral ferromagnetic domain walls \cite{emor} are induced by DM interaction. In contrast to these materials where the ratio of the DM interaction to the exchange interaction couplings are conventionally fixed and relatively small, the DM interaction here can be tuned up to the order of the exchange interaction energy by adjusting the intensity or polarization of the Raman lasers \cite{lin1, wang, cheu, huang}.

\section{Various spin ground state} 
\label{SecIV}

For symmetry reasons and taking the implication of experiments into account, we restrict our discussion in the regime of $\kappa \in \left[0, ~ \frac{\pi}{2} \right]$ and $\lambda \in \left[\frac{1}{2},~2\right]$. Since quantum calculations for Eq.(\ref{MI_eq1}) is nearly intractable, we investigate its magnetic phase within mean-field regime, where a quantum spin is treated into a classical one. The relevant phase diagram is obtained by classical Monte Carlo method. Before performing that, we start our discussion from some extreme cases and obtain some analytic results and generic sense, which serve for further understanding in more complex and general cases. For example, for $\Omega = 0$, the negative $J_{\bot}^{\alpha}$ indicates that spin structure in one layer is a copy of the other layer and thus the issue of double layers can be simplified into the single layer under the treatment of mean field. Similarly the negative $J_{\hat{y}}^{\alpha}$ makes the 2D issue into 1D issue. In what following, we begin to discuss magnetic properties in the extreme case of SO coupling $\Omega \to 0$ and then transit to general cases with finitely large SO coupling.

   \subsection{In the limit of $\Omega \to 0$}

      For a finitely large $J$ and $\Omega \to 0$, the effective Hamiltonian can be easily derived from Eq.(\ref{MI_eq1}) by the meanfield approximation mentioned above, 
      \bea
         H_{\rm spin} = & - & \frac{4 t^{2}}{U} \sum_{i} \left[ \left(2 - \frac{1}{\lambda} \right) S_{i}^{z} S_{i + \hat{x}}^{z} + \frac{\cos(2 \kappa)} {\lambda} \left( S_{i}^{x} S_{i + \hat{x}}^{x}  \right.\right. \nn
                        & + & \left.\left. S_{i}^{y} S_{i + \hat{x}}^{y} \right) + \frac{\sin(2 \kappa)} {\lambda} \left( S_{i}^{x} S_{i + \hat{x}}^{y} - S_{i}^{y} S_{i + \hat{x}}^{x} \right) \right],   \label{Hamiltonian7}
      \ea
which is the combination of the standard 1D $XXZ$ model and DM interaction. 

In absence of DM term ($\kappa = 0$), for $\lambda < 1$, $H_{\rm spin}$ has a $XY$ paramagnetic ground state, while for $\lambda > 1$, it displays a ferromagnet along the $\hat{z}$ direction. We now capture some solvable cases for $\kappa \neq 0$, which are partially similar to Ref.\cite{xu}.       
      
      For $\lambda = 1$, one can perform rotation transformation for each spin $S_{i}^{\alpha}$ along its $S_{i}^{z}$ axis, then choose a constant rotation angle difference between the two nearest-neighbor spin operators, and finally obtain an isotropic ferromagnetic Heisenberg model \cite{perk}. Specifically, define $\tilde{S}_{i}^{\dag} \equiv \exp(-i \theta_{i}S_{i}^{z}) S_{i}^{+} \exp(i \theta_{i} S_{i}^{z}) = \exp(-i \theta_{i}) S_{i}^{+}$, where $S_{i}^{+} = S_{i}^{x} + iS_{i}^{y}$ is the spin-raising operator. Such rotation transformation does not change spin in $\hat{z}$ direction, i.e., $\tilde{S}_{i}^{z} = S_{i}^{z}$. By choosing $\theta_{i} - \theta_{i+1} = 2 \kappa$, one can obtain $H_{\rm spin} = \frac{- 4 t^{2}}{U} \sum_{i, \alpha = x, y , z} \tilde{S}_{i}^{\alpha} \tilde{S}_{i + \hat{x}}^{\alpha}$. The ground state of such Hamiltonian is a ferromagnet and its elementary excitations are spin wave with quadratic dispersion. Explicitly, the original spin states are a spiral state with wave vector $2\kappa$ along $\hat{x}$ direction.

      For $\lambda = \frac{1}{2}$ and in the treatment similar to $\lambda = 1$, the $z$-component of each spin disappear and Hamiltonian (\ref{Hamiltonian7}) transits into a  XY model. This indicates that the spin states are spirals around the $\hat{z}$ axis along the chain. Whilst for $\lambda \rightarrow \infty$, the Hamiltonian becomes the ferromagnetic Ising model. 
      
      When $\kappa = \frac{\pi}{4}$, $H_{\rm spin} = - \frac{4t^{2}}{U} \sum_{i} [(2 - \frac{1}{\lambda}) S_{i}^{z} S_{i + \hat{x}}^{z} + \frac{1}{\lambda} ( S_{i}^{x} S_{i + \hat{x}}^{y} - S_{i}^{y} S_{i + \hat{x}}^{x} ) ]$. This is the combination of the one-dimensional Ising model and DM interaction and has been studied by the quantum renormalization-group and exact diagonalization methods \cite{jafa}. For $\lambda < 1$, the DM term dominates and the system is a chiral $xy$ magnet, while for $\lambda > 1$, it is a $\hat{z}$-direction ferromagnetic state.

   \subsection{Spin-orbit coupling $\Omega \neq 0$}
   
   For finitely large SO coupling, the stagger ``magnetic field'' in the bilayer system competes with the interlayer ferromagnetic coupling, which is contributed by the laser-assisted tunneling $J$. This results in possible complex magnetic structure, which is dependent on the layers. Besides when $\Omega$ becomes sufficiently large, the ``magnetic field'' can break inlayer magnetic structures, e.g., chiral magnet, so that the mechanism of magnetization with respect to different phases becomes not explicit.

   For $\kappa = 0$, the DM term is absent and the system is inlayer ferromagnet. Yet there exists the competition between ``magnetic field'' raised by SO coupling and ferromagnetic Heisenberg term. Since the interlayer spin-spin interaction does not affect such competition, such case can be interpreted by the model of the two-site antiferromagnetic Heisenberg in a magnetic field, specifically, $H_{\rm spin} = \frac{4J^{2}}{U\lambda} (\tilde{S}_{1}^{x} \tilde{S}_{2}^{x} + \tilde{S}_{1}^{y} \tilde{S}_{2}^{y}) + 2 \Omega (\tilde{S}_{1}^{y} + \tilde{S}_{2}^{y})$, where $\tilde{S}_{1}^{\alpha} = - S_{1}^{\alpha}$ and $\tilde{S}_{2}^{\alpha} = S_{2}^{\alpha}$. In a classical treatment, it is obtained that when $\Omega U \lambda \leq 4 J^{2}$, the angle between two spins is $2\arcsin(\Omega U \lambda/4J^{2})$ and $\pi$ for $\Omega U \lambda > 4 J^{2}$.    
      
   \begin{figure}[bt]
      \centering
      \includegraphics[scale = 0.36]{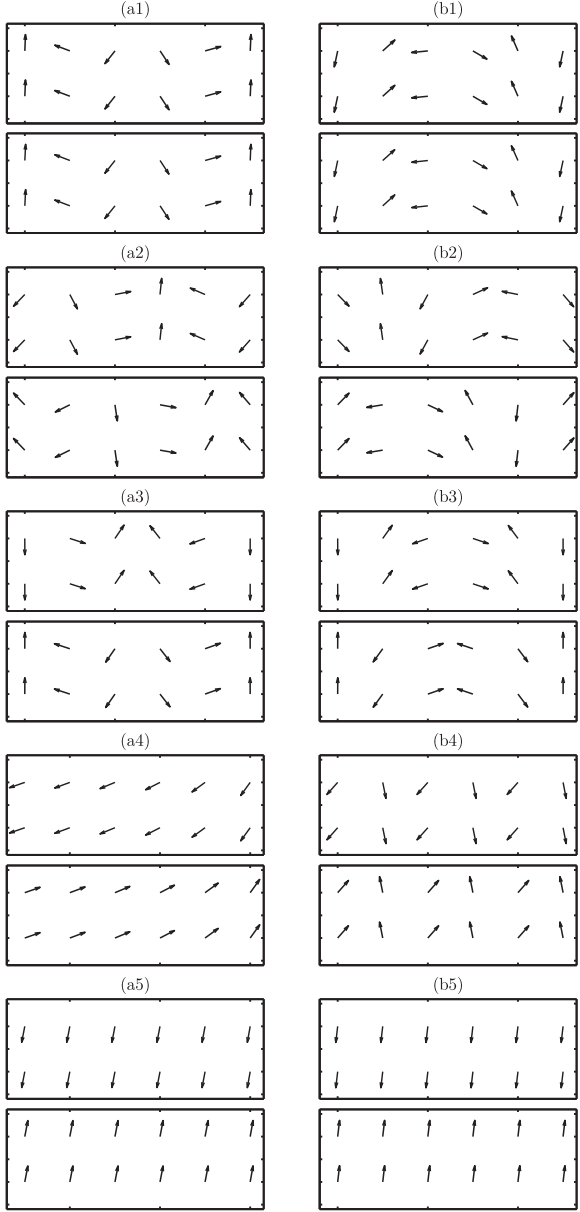} 		        	
      \caption{Spin configurations for upper layer ($a = 2$) and lower layer ($a = 1$) shown in the upper plot and lower plot in each panel, respectively. We display the influence of an increasing strength of spin-orbit coupling on two representative chiral states with respect to $\kappa = 0.2 \pi$ [(a1)-(a5)] and $\kappa = 0.4\pi$ [(b1)-(b5)]: $\tilde{\Omega} = 0$ [(a1) and (b1)]; $\tilde{\Omega} = 0.75$ [(a2) and (b2)]; $\tilde{\Omega} = 1.25$ [(a3) and (b3)]; $\tilde{\Omega} = 50$ [(a4)] and $75$ [(b4)]; $\tilde{\Omega} = 250$ [(a5) and (b5)]. Other parameters are $\tilde{t} = 10$ and $\lambda = 0.5$.}
      \label{state}
   \end{figure}      

More generic cases for $\kappa \neq 0$ are conventionally analytically inaccessible, so we investigate the associated magnetic properties by numerically solving $H_{\rm spin}$ in Eq.(\ref{MI_eq1}) via Monte Carlo annealing. For dimensionless parameters, we choose $\frac{4J^{2}}{U\lambda}$ as characteristic energy unit and define $\tilde{\Omega} = \frac{\Omega U \lambda}{4 J^{2}}$, $\tilde{t} = \frac{t}{J}$, and $\tilde{E}_{\rm gs} = \frac{E_{\rm gs} U \lambda}{4 J^{2}}$ where $E_{\rm gs}$ is the ground-state energy averaged over the number of sites. Since the potential physics for $\lambda > 1$ is relatively explicit, we concentrate on much more complex cases with $\lambda < 1$. Without loss of generality, we choose $\lambda = \frac{1}{2}$ in the following part and thus the $z$-component of each spin is zero.  

To identify the various magnetic phase, we define the average interlayer phase difference $\Gamma = \frac{1}{M} \sum_{i, j} |\theta_{i, j, 2} - \theta_{i, j, 1}|$, where $\theta_{i, j, a}$ is the angle of spin in the spatial grid position $(i, j)$ of $a$-th layer and $M$ is the total number of counted spins. Also the chiral order of spatial period is denoted by $\mathcal{D}$, which is arisen from the period of sine function in Eq.(\ref{MI_eq1}). From the above derivation for $\lambda = 1$ and $\Omega \to 0$, $\mathcal{D}$ is  the \emph{minimal} positive integer which fulfills $\mathcal{D} = \frac{\pi n}{\kappa}$, where $n \in \mathbb{Z}_{\geq 0}$. 

Fig.\,\ref{state} shows the influence of an increasing strength of spin-orbit coupling $\Omega$ on two representative chiral states. As it is stressed, we perform many calculations according to a variety of system size and find that results are qualitatively identical. For $\Omega = 0$, the finitely large interlayer tunneling $J$ makes the spin structure in one layer as a copy of the other layer [see (a1) and (b1)]. We denote such phase as chiral phase with interlayer ferromagnet (IFM). With a slightly increasing $\Omega$, there is an angle between any two spins in the same position but different layer, $\theta_{i, j, 2} - \theta_{i, j, 1}$, which varies from 0 to $\pi$ [see (a2) and (b2)]. To the best of our knowledge, two chiral states with a fixed angle between any two spins in different layer/film occur extremely rarely in the region of material science. We label such phase with $0<\Gamma<\pi$ as heterochiral phase. Furthermore, when $\Omega$ is larger than a critical value, there exists antiferromagnetic correlation between any two spins $S_{i, j, 2}$ and $S_{i, j, 1}$ [see (a3) and (b3)]. We find that the critical value $\Omega_{c} = \frac{4 J^{2}}{U\lambda}$, which originates from the model of two-site antiferromagnetic Heisenberg in a magnetic field. Note that up to the critical value, the chiral order is not broken, i.e., $\mathcal{D} = 5$ corresponding to $\kappa = 0.2\pi$ and $0.4\pi$ remains. We call such phase as chiral phase with interlayer antiferromagnet(IAFM). However, with a large increasing $\Omega$, the ``magnetic field'' term affects significantly the inlayer chiral configuration, and the mechanism of magnetization for $\kappa = 0.2 \pi$, where the nearest neighbor sites in $\hat{x}$ direction have ferromagnetic correlation, and for $\kappa = 0.4 \pi$, where it displays the antiferromagnet, are different. In generic, for $\kappa = 0.2 \pi$, $\mathcal{D}$ prefers to increase before the chiral state is completely magnetized (such state is referred as FM with IAFM) [see (a4) and (a5)], while for $\kappa = 0.4 \pi$, $\mathcal{D}$ varies toward $\mathcal{D} = 2$ prior to the occurrence of FM with IAFM [see (b4) and (b5)]. Here we denote the state with $\mathcal{D} = 2$ as stripe state, a special chiral state. We argue that the distinct mechanism of magnetization for two types of chiral phase stems from the different spin correlation in the nearest neighbor sites, e.g., ferromagnet for $\kappa = 0.2\pi$ and antiferromagnet for $\kappa = 0.4\pi$ along $\hat{x}$ direction.

\begin{figure}[tb]
\centering
\includegraphics[scale = 0.345]{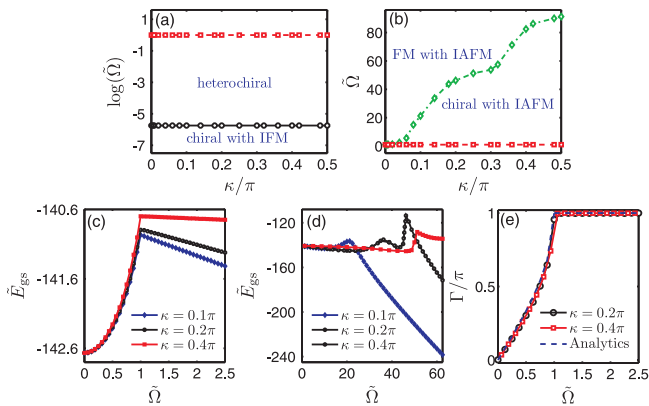} 		        	
\caption{Phase diagram and associated physics qualities from Monte Carlo simulation of the spin Hamiltonian (\ref{MI_eq1}) for $\tilde{t} = 10$ and $\lambda = 0.5$. (a) shows the part of phase diagram in the regime of small spin-orbit coupling, i.e, $\tilde{\Omega} \leq 1$ and (b) the part for large $\tilde{\Omega}$. (c) and (d) shows the average ground-state energy for small $\tilde{\Omega}$ and large $\tilde{\Omega}$, respectively. (e) provides the average interlayer phase difference with respect to different $\kappa$. The blue dashed line in (e) is the numerical evaluation of the analytic function for the angle of two spins in the model of the two-site antiferromagnetic Heisenberg in a magnetic field, i.e., $2 \arcsin(\Omega U \lambda /4 J^{2})$ for $\Omega U \lambda \leq 4 J^{2}$ and $\pi$ otherwise. Markers are numerically evaluated data points, and lines are a guide to the eye.}
\label{phase}
\end{figure}

On basis of intensive calculations, we summarize a representative phase diagram for such bilayer system and show it in Fig.\,\ref{phase}. The phase boundary is identified by comparing unbiased Monte Carlo annealed energies and variational energies of a variety of states at various $\kappa$ as well as $\Omega$ \cite{cole}. Also we determine the boundary by analyzing the associated ground-state spin configuration via some physical qualities, such as $\Gamma$ and $\mathcal{D}$. Moreover, $\kappa = 0$ and $\kappa = 0.5\pi$ correspond to inlayer FM and AFM along $\hat{x}$ direction, respectively. Here we do not provide special demonstration for such two special cases in Fig.\,\ref{phase}, because FM and AFM can be essentially taken as a special chiral phase, i.e., $\mathcal{D} = 1$ for FM and 2 for AFM. The transition between chiral with IFM and heterochiral as well as between heterochiral and chiral with IAFM (see Fig.\,\ref{phase} (a) and (b)) are attributed from the competition between SO coupling and laser-assisted tunneling. It can be explained approximately by the model of the two-site antiferromagnetic Heisenberg in a magnetic field. The energy variation from chiral with IFM to heterochiral is nearly smooth [see Fig.\,\ref{phase} (c)]. We distinguish two types of phases by identifying its spin configuration: the state with $\Gamma < \Gamma_{c}$ where $\Gamma_{c}$ is a small value refers to chiral with IFM and the state with $\Gamma_{c} \leq \Gamma <\pi$ corresponds to heterochiral.    
In our parameter regime, the transition point occurs in $\frac{\Omega U \lambda}{4 J^{2}} \approx 0.003$ where $\Gamma_{c} \approx 10^{-3}\pi$. By contrast, the energy variations corresponding to $\Omega$ in the region of heterochiral and chiral with IAFM are explicitly different; the tunneling point between two phases appears exactly in $\frac{\Omega U \lambda}{4 J^{2}} = 1$. It is manifested that the parameter regime for heterochiral can be significantly enlarged for a large $J$. Also we note that the average interlayer phase difference in these transitions can be described well by the model of two-site antiferromagnetic Heisenberg in a magnetic field [see Fig.\,\ref{phase} (e)]. For a larger $\Omega$, the stagger ``magnetic field'' can be comparable with the tunneling term between lattices $t$. The corresponding chiral order of spatial period becomes sensitive for $\Omega$ as well as $\kappa$, so that the transition point between chiral with IAFM and FM with IAFM are different with respect to various $\kappa$ [see Fig.\,\ref{phase} (b) and (d)]. In generic, a larger $\kappa$ results in a larger $\Omega_{c}$. After complete magnetization, no more phase occurs.

\section{Discussion and Conclusion} \label{secV}

The critical part for experimental implementation of the proposed bilayer SO-coupled system is to generate Raman-induced SO coupling and laser-assisted interlayer tunneling. The details for such generation is shown in Appendix \ref{ALI}. Two magnetic sublevels of the $^{87}$Rb-type alkali atoms can be a candidate for the atomic internal (quasi-$1/2$) states, e.g., $|F = 1, m_{F} = 0\rangle \equiv |\uparrow\rangle$ and $|F = 1, m_{F} = -1\rangle \equiv |\downarrow\rangle$. Since the magnetic phases in our bilayer system is qualitatively identical for $\lambda < 1$ (as well as for $\lambda > 1$), it indicates that there is a large degree of freedom for experiments to adjust the parameters of interaction in optical lattice according to the variant of spin configurations. A large challenge for the bilayer system is to measure its spin correlation for the complex magnetic phase, especially for chiral magnetic order. We suggest experiments to measure the case of $\Omega \to 0$, which corresponds to interlayer FM, in the first step, and then probe for the case of a relatively large $\Omega$, which matches interlayer AFM. The experimental techniques for observing FM and AFM in cold atoms have existed, for example, spin-sensitive Bragg scattering experiments \cite{hart, coro} and \textit{in situ} microscopy \cite{weit, sher, bakr}. After two measurements for interlayer FM and AFM, experiments can explore the generic case of moderate $\Omega$ and compare its results with the above two cases. 

In summary, we have found a variety of magnetic phases in the bilayer Bose-Hubbard model with 1D SO coupling.
The representative magnetic phases, such as hetero ferromagnet and heterochiral magnet, are indicative for the competition between SO coupling and laser-assisted interlayer tunneling. We also note that in the realm of solid-state materials, the heterochiral magnet induced by SO coupling occurs extremely rarely. Our results offer a theoretical understanding for complex magnetic phenomena, which is involved with interlayer tunneling and SO coupling, and thus inspire material science to realize such magnetic phase. Moreover, in contract to the chiral phase with the ferromagnetic correlation between the nearest neighbor sites and the chiral phase with the antiferromagnetic correlation along $\hat{x}$ direction, the strong `` magnetic field'' arisen from SO coupling breaks the chiral order of spatial period, $\mathcal{D}$, in different ways: in the former, $\mathcal{D}$ prefers to increase before it is fully magnetized while decrease to $\mathcal{D} = 2$ (stripe state) in the latter. The experimental realization for such bilayer model has been intensively discussed.

\acknowledgments
We thank J.-S. You and R.-C. Lin for useful discussions. The work was supported by NCTS and MoST in Taiwan. 

\appendix

\section{Atom-Light Interaction} \label{ALI}

The laser-induced transition between two internal state $H_{\rm intra}^{'}$ and the laser-assisted intralayer tunneling $H_{\rm inter}^{'}$ are essentially originated from the coupling between atoms and laser. Specifically, a general Hamiltonian for the atom-light interaction in an atomic hyperfine ground state according to the total spin $F$ has the form \cite{gold},
   \bea
     H_{\rm AL} = u_{s} ({\bf E}^{*} \cdot {\bf E}) + \frac{i u_{v} g_{F}}{\hbar g_{J}} ({\bf E}^{*} \times {\bf E} ) \cdot \hat{\bf F},
   \ea
where ${\bf E}$ is the negative-frequency part of the full electronic field. $u_{s}$ and $u_{v}$ are the scalar and vector atomic polarizabilities. The tensor polarizability is neglibible for the case of the detuning of the off-resonance light field much larger than the fine-structure splitting of the excited electronic state. $g_{J}$ and $g_{F}$ are the hyperfine Land{\' e} $g$-factors for the electronic spin and the total angular momentum of the atom, respectively. As proposed in the main text, $^{87}$Rb atoms with two magnetic sublevels, i.e., $|F = 1, m_{F} = 0 \rangle \equiv |\uparrow \rangle$ and $|F = 1, m_{F} = -1\rangle \equiv |\downarrow \rangle$, can be chosen for the atomic internal states in the bilayer system and the relevant $g_{F} / g_{J} = -1/4$. To avoid the involvement of the state $|F = 1, m_{F} = 1 \rangle$ in the ring coupling scheme, a sufficiently large magnetic field is required to generate a quadratic Zeeman shift as a result that the detuning between $|F = 1, m_{F} = 1 \rangle$ and $|\uparrow \rangle$ is much larger than the detuning between $|\uparrow \rangle$ and $|\downarrow \rangle$ (denoted by $\Delta_{\rm intra}$). Additionally, to employ the minimum number of lasers in the scheme, a spin-independent asymmetric double-well potential is preferred and the bias energy for the atomic ground states localized in two layers, $\Delta_{\rm inter}$, can be made much smaller than the line width of electronic excited states.

Figure \ref{Fig1} (a) illustrates a specific laser configuration which customizes the desirable intra- and interlayer couplings shown in Fig.\,\ref{Fig1} (b). Here both layers are simultaneously illuminated by three laser beams labeled by ${\bf E}_{0}$, ${\bf E}_{1}$, and ${\bf E}_{2}$. Note that this experimentally accessed setup distinguishes from the one proposed by \cite{sun}, which is extremely difficult to be realized experimentally: the Raman transitions should be accompanied by a recoil in different direction for different layers and the interlayer laser-assisted tunneling should be realized by a recoil in different direction for different spin states. 

We propose three lasers in such means, ${\bf E}_{0} \sim \hat{\bf x} e^{i (k_{0} z - \omega_{0} t)}$, ${\bf E}_{1} \sim \hat{\bf y} e^{i \left[k_{x} x - (\omega_{0} + \delta \omega_{1}) t\right]}$, ${\bf E}_{2} \sim \hat{\bf x} e^{i \left[k_{y} y - (\omega_{0} + \delta \omega_{2}) t\right]}$. 
Explicitly, ${\bf E}_{0}^{*} \cdot {\bf E}_{1} = 0$ while ${\bf E}_{0}^{*} \times {\bf E}_{1}$ is nonzero. This indicates that the intralayer spin-flip transitions emerge when the frequencies of the fields ${\bf E}_{0}$ and ${\bf E}_{1}$ are tuned to the two-photon resonance, i.e., $\delta \omega_{1} = \Delta_{\rm intra}$. Therefore, the Hamiltonian of the intralayer Raman coupling can be written as 
   \bea
      \hat{H}_{\rm intra}^{'} = && \int dx dy \sum_{a = 1, 2} \left\{ \Omega e^{i \left[k_{x} x + (-1)^{a} \phi - \delta \omega_{1} t \right]} + {\rm c.c.} \right\}  \nn
                               && \hat{\psi}_{a\uparrow}^{\dag}({\bf r}) \hat{\psi}_{a\downarrow}({\bf r}) + {\rm h.c.} \label{Hintra}
   \ea
Due to the strong confinement of double well along $\hat{z}$ direction, the out-of-plane Raman recoil provides the difference $2\phi = k_{0} d$ for the Raman coupling in different layers. Clearly the phase difference can be tuned by either varying the double-well separation $d$ or the out-of-plane Raman recoil $k_{0}$. As we concentrate on, $\phi = \pi/2$ to fulfill the ring tunneling scheme \cite{camp}. 

The interlayer tunneling is attributed by ${\bf E}_{0}$ and ${\bf E}_{2}$, where ${\bf E}_{0}^{*} \cdot {\bf E}_{2} \neq 0$ while ${\bf E}_{0}^{*} \times {\bf E}_{2} = 0$. To achieve the state-independent interlayer transition, the oscillation frequency of a scalar light shift $\delta \omega_{2}$ is equal to $\Delta_{\rm inter}$. The resulting Hamiltonian for the laser-assisted tunneling yields, 
   \bea
      \hat{H}_{\rm inter}^{'} = && \int\int dx dy \sum_{\gamma = \uparrow, \downarrow} \left[ J e^{i(k_{y}y - \delta \omega_{2} t)} + {\rm c.c.} \right] \nn
                                &&\hat{\psi}_{\gamma, 2}^{\dag} ({\bf r}) \hat{\psi}_{\gamma, 1} ({\bf r}) + {\rm h.c.},   \label{Hinter}
   \ea  
where $J = \Omega_{J} \int dz \psi_{2}^{*} (z) \psi_{1}(z) e^{i k_{0} z}$ is the interlayer coupling. $\Omega_{J}$ is the associated Rabi frequency and $\psi_{1, 2} (z)$ are the Wannier-like states localized at layer 1 or 2. Since $\psi_{1}(z)$ and $\psi_{2} (z)$ are orthogonal, the non-vanishing overlap integral $J$ is contributed by the factor $e^{i k_{0} z}$. 

Finally, the time-dependent terms in $\hat{H}_{\rm intra}^{'}$ and $\hat{H}_{\rm inter}^{'}$ can be removed in the rotating frame. Also without loss of generality, we consider the length of the in-plane wave vectors are nearly the same, i.e., $k_{x} = k_{y} = \kappa$.

\section{Single-Particle Dispersion} \label{SPD}

\begin{figure}[bt]
\centering
\includegraphics[scale = 0.44]{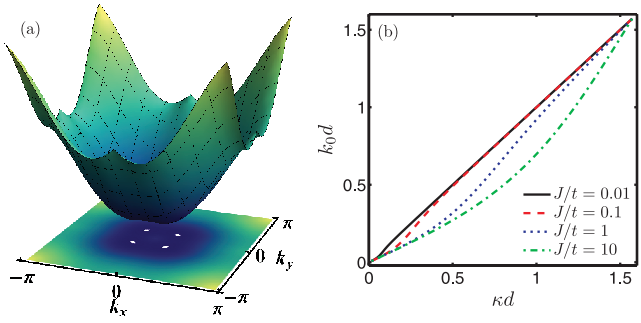} 		        	
\caption{(Color online) (a) The lowest energy spectrum, $E_{-, +}(k_{x}, k_{y})$ for $t = 1$, $\Omega = J = 2$, and $\kappa = \pi/4$. (b) The minimum wave vector $k_{0}$ with respect to the wave vector of the lasers, $\kappa$. Different plots correspond to various $J$ ( $= \Omega$). 
}
\label{Fig3}
\end{figure}

By performing Fourier transformation for Eq.(\ref{hamiltonian3}), we write out the Hamiltonian $H_{\bf k}$ in momentum space and after diagonalizing $H_{\bf k}$, we obtain four branches of energy spectra, 
   \begin{widetext} 
      \begin{equation} \label{Hamiltonian_eq2}
         E_{\pm, \pm} (k_{x}, k_{y})  =  -2t (\cos k_{x} + \cos k_{y}) \cos \kappa \pm \sqrt{4t^{2} (\sin^{2} k_{x} + \sin^{2} k_{y}) \sin^{2} \kappa + (\Omega^{2} + J^{2}) \pm 4t \sin\kappa \sqrt{\mathcal{A}} },
      \end{equation}
   \end{widetext}
where $ \mathcal{A} = 4t^{2} \sin^{2} k_{x} \sin^{2} k_{y} \sin^{2} \kappa + \Omega^{2} \sin^{2} k_{y} + J^{2} \sin^{2} k_{x}$. Supposing $\kappa \in \left[0,~ \frac{\pi}{2} \right]$ (note $d = 1$) and because of $t > 0$, the lowest energy band fulfills $E_{-, +} (k_{x}, k_{y})$. In generic, there exist four degenerate minima in the lowest band at $\vec{Q}_{1} = (k_{0}, k_{0}^{'})$, $\vec{Q}_{2} = (-k_{0}, k_{0}^{'})$, $\vec{Q}_{3} = (-k_{0}, -k_{0}^{'})$, and $\vec{Q}_{4} = (k_{0}, -k_{0}^{'})$, as shown in Fig.\,\ref{Fig3} (a). 

For $\Omega = J$, the minima prefer $k_{0} = k_{0}^{'}$ so that the ground-state energy is similar to the one in the system of 2D OL with Rashbar SO coupling where the lower band has the form, $E_{\rm OL} = - 2t (\cos k_{x} + \cos k_{y}) \cos \beta - 2t \sin\beta\sqrt{\sin^{2} k_{x} + \sin^{2} k_{y}}$ with $\beta$ Rashbar SO coupling length \cite{cole}. It is stressed that in the 2D OL, there exists a straight relation between the ground-state momentum and SO coupling strength, i.e., $\tan k_{0} = (\tan \beta)/\sqrt{2}$, while the single relation between $k_{0}$ and $\kappa$ is absent in our bilayer system. Fig.\,\ref{Fig3} (b) shows the typical relation between $k_{0}$ and $\kappa$ in the case of $\Omega = J$.  For $t \gg \Omega, J$, $E_{-, +} (k_{x}, k_{y}) \approx - 2t (\cos k_{x} + \cos k_{y}) \cos \kappa - 2t \sin \kappa (|\sin k_{x}| + |\sin k_{y}|)$, and through $\partial E_{-, +} /\partial k_{x} = \partial E_{-, +} /\partial k_{y} = 0$, one can find $k_{0} = \kappa$. This matches the dark solid line in Fig.,\ref{Fig3} (b) where $J / t = 0.01$. For $t \ll \Omega, J$, it can be obtained that $\tan k_{0} = \tan \kappa /2$, which coincides to the green line in Fig.,\ref{Fig3} (b) where $J/ t = 100$.  

For $\Omega \neq J$, the ground state favors $k_{0} \neq k_{0}^{'}$. Since $\Omega$ and $J$ play similar role in energy spectrum, according to various $\kappa$, the case of $\Omega \neq J$ affect merely the ratio between $k_{0}$ and $k_{0}^{'}$ but don't show essential difference (or tendency) between them.



\begin{thebibliography}{000}

   \bibitem{lin1} Y.-J. Lin, K. Jim\`{e}nez-Garc\'{i}a, and I. B. Spielman, \href{http://dx.doi.org/10.1038/nature09887}{ Nature(London), {\bf 471}, 83 (2011)}.
   
   \bibitem{wang} P. Wang, Z.-Q. Yu, Z. Fu, J. Miao, L. Huang, S. Chai, H. Zhai, and J. Zhang, \href{http://dx.doi.org/10.1103/PhysRevLett.109.095301} {Phys. Rev. Lett. {\bf 109}, 095301 (2012)}.
   
   \bibitem{cheu} L. W. Cheuk, A. T. Sommer, Z. Hadzibabic, T. Yefsah, W. S. Bakr, and M. W. Zwierlein, \href{http://dx.doi.org/10.1103/PhysRevLett.109.095302} {Phys. Rev. Lett. {\bf 109}, 095302 (2012)}. 

   \bibitem{radi} J. Radi{\' c}, A. D. Ciolo, K. Sun, and V. Galitski, \href{http://dx.doi.org/10.1103/PhysRevLett.109.085303} {Phys. Rev. Lett. {\bf 109}, 085303 (2012)}. 
   
   \bibitem{cole} W. S. Cole, S. Z. Zhang, A. Paramekanti, and N. Trivedi, \href{http://dx.doi.org/10.1103/PhysRevLett.109.085302} {Phys. Rev. Lett. {\bf 109}, 085302 (2012)}. 
   
   \bibitem{sino} J. Sinova, D. Culcer, Q. Niu, N. A. Sinitsyn, T. Jungwirth, and A. H. MacDonald, \href{http://dx.doi.org/10.1103/PhysRevLett.92.126603} {Phys. Rev. Lett. {\bf 92}, 126603 (2004)}. 

   \bibitem{kane} C. L. Kane and E. J. Mele, \href{http://dx.doi.org/10.1103/PhysRevLett.95.146802}{Phys. Rev. Lett. {\bf 95}, 146802 (2005)}. 

   \bibitem{ruse} J. Ruseckas, G. Juzeli$\bar{\rm u}$nas, P. {\" O}hberg, and M. Fleischhauer, \href{http://dx.doi.org/10.1103/PhysRevLett.95.010404} {Phys. Rev. Lett.  {\bf 95}, 010404 (2005)}. 
   
   \bibitem{juze1} G. Juzeli$\bar{\rm u}$nas, J. Ruseckas, M. Lindberg, L. Santos, and P. {\" O}hberg, \href{http://dx.doi.org/10.1103/PhysRevA.77.011802} {Phys. Rev. A {\bf 77}, 011802(R) (2008)}. 

   \bibitem{juze2} G. Juzeli$\bar{\rm u}$nas, J. Ruseckas, and J. Dalibard, \href{http://dx.doi.org/10.1103/PhysRevA.81.053403} {Phys. Rev A {\bf 81}, 053403 (2010)}. 
   
   \bibitem{camp} D. L. Campbell, G. Juzeli$\bar{\rm u}$nas, and I. B. Spielman, \href{http://dx.doi.org/10.1103/PhysRevA.84.025602} {Phys. Rev. A {\bf 84}, 025602 (2011)}. 
   
   \bibitem{liu} X.-J. Liu, K. T. Law, and T. K. Ng, \href{http://dx.doi.org/10.1103/PhysRevLett.112.086401} {Phys. Rev. Lett. {\bf 112}, 086401 (2014)}.

   \bibitem{ande} B. M. Anderson, I. B. Spielman, and G. Juzeli$\bar{\rm u}$nas, \href{http://dx.doi.org/10.1103/PhysRevLett.111.125301} {Phys. Rev. Lett. {\bf 111}, 125301 (2013)}.
   
   \bibitem{xu1} Z.-F. Xu, L. You, and M. Ueda, \href{http://dx.doi.org/10.1103/PhysRevA.87.063634} {Phys. Rev. A {\bf 87}, 063634 (2013)}. 

   \bibitem{xu2} Z.-F. Xu and L. You, \href{http://dx.doi.org/10.1103/PhysRevA.85.043605} {Phys. Rev. A {\bf 85}, 043605 (2012)}. 

   \bibitem{sun} Q. Sun, L. Wen, W.-M. Liu, G. Juzeli$\bar{\rm u}$nas, and A.-C. Ji, \href{http://dx.doi.org/10.1103/PhysRevA.91.033619}{Phys. Rev. A {\bf 91}, 033619 (2015)}.   

   \bibitem{huang} L. H. Huang, Z. M. Meng, P. J. Wang, P. P. S.-L. Zhang, L. C. Chen, D. H. Li, Q. Zhou, and J. Zhang, \href{http://dx.doi.org/10.1038/NPHYS3672}{Nature Phys. {\bf 12}, 540 (2016)}. 
   
   \bibitem{note1} The experimental setup of the bilayer system is provided in Appendix A and the derivation of 2D SO coupling characterized by the single-particle dispersion is shown in Appendix B. 



   \bibitem{lin3} Y.-J. Lin, R. L. Compton, K. Jim\`{e}nez-Garc\'{i}a, J. V. Porto, and I. B. Spielman, \href{http://dx.doi.org/10.1038/nature08609}{Nature(London), {\bf 462}, 628 (2009)}. 

   \bibitem{hart} R. A. Hart, P. M. Duarte, T.-L. Yang, X. Liu, T. Paiva, E. Khatami, R. T. Scalettar, N. Trivedi, D. A. Huse, and R. G. Hulet, \href{http://dx.doi.org/10.1038/nature14223}{Nature {\bf 519}, 211 (2015)}.   
   
   \bibitem{coro} T. A. Corovilos, S. K. Baur, J. M. Hitchcock, E. J. Mueller, and R. G. Hulet, \href{http://dx.doi.org/10.1103/PhysRevA.81.013415}{Phys. Rev. A {\bf 81}, 013415 (2010)}.
    
   \bibitem{gold} N. Goldman, G. Juzeli$\bar{\rm u}$nas, P. {\"{O}}hberg, and I. B. Spielman,  \href{http://dx.doi.org/10.1088/0034-4885/77/12/126401}{Rep. Prog. Phys. {\bf 77}, 126401 (2014)}.
   
   \bibitem{su} S.-W. Su, S.-C. Gou, Q. Sun, L. Wen, W.-M. Liu, A.-C. Ji, J. Ruseckas, and G. Juzeli$\bar{\rm u}$nas, \href{http://dx.doi.org/10.1103/PhysRevA.93.053630}{Phys. Rev. A {\bf 93}, 053630 (2016)}.     
  
   \bibitem{dzya} I. Dzyaloshinskii, \href{http://dx.doi.org/10.1016/0022-3697(58)90076-3}{J. Phys. Chem. Solids {\bf 4}, 241 (1958)}. 
   
   \bibitem{mori} T. Moriya, \href{http://dx.doi.org/10.1103/PhysRev.120.91}{Phys. Rev. {\bf 120}, 91 (1960)}. 

   \bibitem{bode} M. Bode, M. Heide, K. von Bergmann, P. Ferriani, S. Heinze, G. Bihlmayer, A. Kubetzka, O. Pietzsch, S. Bl{\" U}gel, and R. Wiesendanger, \href{http://dx.doi.org/10.1038/nature05802}{Nature {\bf 447}, 190 (2007)}.

   \bibitem{heid} M. Heide, G. Bihlmayer, and S. Bl{\" U}gel, \href{http://dx.doi.org/10.1103/PhysRevB.78.140403}{Phys. Rev. B {\bf 78}, 140403(R)(2008)}.

   \bibitem{yu} X. Z. Yu, Y. Onose, N. Kanazawa, J. H. Park, J. H. Han, Y. Matsui, N. Nagaosa, and Y. Tokura, \href{http://dx.doi.org/10.1038/nature09124}{Nature, {\bf 465}, 901 (2010)}.  
   
   \bibitem{hein} S. Heinze, K. von Bergmann, M. Menzel, J. Brede, A. Kubetzka, R. Wiesendanger, G. Bihlmayer, and S. Bl{\" u}gel, \href{http://dx.doi.org/10.1038/nphys2045}{Nature Phys. {\bf 7}, 713 (2011)}.   

   \bibitem{huang1} S. X. Huang  and C. L. Chien, \href{http://dx.doi.org/10.1103/PhysRevLett.108.267201}{Phys. Rev. Lett. {\bf 108}, 267201 (2012)}.
   
   \bibitem{emor} S. Emor, U. Bauer, S.-M. Ahn, E. Martinez, and G. S. D. Beach, \href{http://dx.doi.org/10.1038/nmat3675}{Nature Mater. {\bf 12}, 611 (2013)}.   
   
   \bibitem{xu} Z.-H. Xu, W. S. Cole, and S.-Z. Zhang, \href{http://dx.doi.org/10.1103/PhysRevA.89.051604}{Phys. Rev. A {\bf 89}, 051604 (R) (2014)}.

   \bibitem{perk} J. H. H. Perk and H. W. Capel, \href{http://dx.doi.org/10.1016/0375-9601(76)90515-6}{Phys. Lett. A {\bf 58}, 115 (1976)}.

   \bibitem{jafa} R. Jafari, M. Kargarian, A. Langari, and M. Siahatgar, \href{http://dx.doi.org/10.1103/PhysRevB.78.214414}{Phys. Rev. B {\bf 78}, 214414 (2008)}. 

   \bibitem{note2} Under such consideration, the first- and third-order perturbation terms are, indeed, zero so a higher precise should be up to $\mathcal{O} (t^{4}/U)$.

   \bibitem{weit} C. Weitenberg, M. Endres, J. F. Sherson, M. Cheneau, P. Schau$\beta$, T. Fukuhara, I. Bloch, and S. Kuhr, \href{http://dx.doi.org/10.1038/nature09827}{Nature {\bf 471}, 319 (2011)}. 

   \bibitem{sher} J. F. Sherson, C. Weitenberg, M. Endres, M. Cheneau, I. Bloch and S. Kuhr, \href{http://dx.doi.org/10.1038/nature09378}{Nature {\bf 467}, 68 (2010)}.

   \bibitem{bakr} W. S. Bakr, A. Peng, M. E. Tai, R. Ma, J. Simon, J. I. Gillen, S. F{\" o}lling, L. Pollet, and M. Greiner, \href{http://dx.doi.org/10.1126/science.1192368}{Science {\bf 329}, 547 (2010)}.
   

   
%
%
%
%
%

\end{thebibliography}
\end{document}